\documentclass[showpacs,preprint,amsmath,PRB]{revtex4}
\usepackage{graphicx}
\usepackage{dcolumn}
\usepackage{bm}

\begin{document}

\title{Probing semiclassical magneto-oscillations in the low-field quantum Hall effect}

\author{D. R. Hang$^{a,b,\ast}$, C. F. Huang$^{c}$, K. A. Cheng$^{d}$}

\affiliation{ $^{a}$ Department of Materials and Optoelectronic Science, National Sun Yat-sen University, Kaohsiung 804, Taiwan, R.O.C.} 
\affiliation{ $^{b}$Center for Nanoscience and Nanotechnology, National Sun Yat-sen University, Kaohsiung 804, Taiwan, R.O.C.} 
\affiliation{$^{c}$National Measurement Laboratory, Center for Measurement Standards, Industrial Technology Research Institute, Hsinchu 300, Taiwan, R.O.C.}
\affiliation{$^{d}$Department of Electronic Engineering, Lunghwa University of Science and Technology, Taoyuan 333, Taiwan, R.O.C.}

\date{\today}

\begin{abstract}
The low-field quantum Hall effect is investigated on a two-dimensional electron system in an AlGaAs/GaAs heterostructure. Magneto-oscillations following the semiclassical Shubnikov-de Haas formula are observed even when the emergence of the mobility gap shows the importance of quantum localization effects. Moreover, the Lifshitz-Kosevich formula can survive as the oscillating amplitude becomes large enough for the deviation to the Dingle factor. The crossover from the semiclassical transport to the description of quantum diffusion is discussed. From our study, the difference between the mobility and cyclotron gaps indicates that some electron states away from the Landau-band tails can be responsible for the semiclassical behaviors under low-field Landau quantization. \newline
$^{\ast}$ drhang@faculty.nsysu.edu.tw
\end{abstract}

\maketitle

\section{INTRODUCTION}
Recently, there have been renewed interests towards the study of Landau quantization in two-dimensional electron systems (2DESs) under a perpendicular magnetic field $B$. It is well-known that Landau quantization can modulate the density of states and induce magneto-oscillations periodic in the inverse of $B$ in the longitudinal resistivity $\rho _{xx}$. At low enough $B$ where the spin-splitting is unresolved, such oscillations are expected to follow the Shubnikov-de Haas (SdH) formula such that \cite{Coleridge,SdH}
\begin{eqnarray}
 \rho _{xx} (B,T) = \rho_{xx}(B=0) + \Delta \rho_{xx} cos[ \pi ( \nu -1)]
\end{eqnarray}
with the oscillating amplitude
\begin{eqnarray}
\Delta \rho_{xx} (B,T) \sim (X/sinhX)F(B).
\end{eqnarray}
Here $\rho_{xx}(B=0)$ is the value of $\rho _{xx}$ at $B$=0, $\nu$ represents the filling factor, and the parameter $X = 2 \pi ^{2} k _{B} m ^{\ast} T / \hbar e B$ with $k_{B}$, $e$, $\hbar$, $m ^{\ast}$, and $T$ as the Boltzmann constant, electron charge, reduced Plank constant, electron effective mass, and temperature, respectively. The $T$-dependent factor $X/sinhX$ comes from the Lifshitz-Kosevich (LK) formula \cite{Lifshitz,Leadley}, and the $T$-independent factor $F(B)$ is usually taken in the form embodied by the Dingle plot, i.e. the standard Dingle factor \cite{Coleridge,Piot}
\begin{eqnarray}
F(B) = 4c \rho _{xx}(B=0) exp(- \pi / \mu_{q}B).
\end{eqnarray}
Here $\mu _{q}$ denotes the quantum mobility \cite{Harrang} and c is a numerical coefficient in the order of unity \cite{Chen,Hang,Cho}. The SdH theory has been widely used to determine the effective mass, carrier concentration, and quantum mobility in semiconductor heterostructures \cite{Hang2}. Such a theory, including the LK formula, can be derived semiclassically without considering the quantum localization. \cite{Coleridge} On the other hand, the quantum localization is important to the integer quantum Hall effect (IQHE). \cite{Coleridge2,Kivelson,Prange} Because of the localization effects, only states at centers of Landau bands are extended in the high-field IQHE. \cite{Prange} It is believed that such states are distributed within a very narrow energy range. \cite{Wei} The extended states in different Landau bands are separated by the mobility gaps $\Delta E$. As the Fermi level is situated in the middle between adjacent Landau levels, $\Delta E$ can be evaluated at the minimum points of the longitudinal resistivity $\rho _{xx,min}$ by \cite{Leadley,Klitzing}
\begin{eqnarray}
\rho _{xx,min} \underline{\sim} \rho_{0} exp(-\Delta E/2k _{B}T),
\end{eqnarray}
where the prefactor $\rho_{0}$ is independent of $T$. Second-order phase transitions occur as Fermi energy passes through the extended states with sweeping $B$, and universalities of such transitions have been investigated by studying the IQHE. \cite{Wei,Huo,Burgess} 

Despite the success of the SdH and IQHE theories, more studies are still necessary to understand their applicable ranges. The quantum localization is taken into account in the standard IQHE theory even as $B \rightarrow 0$, but the localization length usually becomes much larger than the realistic sample size with decreasing $B$. \cite{Kivelson,Jiang,Fogler} Therefore, the localization effects diminish at low fields, and alternative mechanisms have been discussed to explain how the IQHE-like properties survive as such effects are reduced. \cite{Coleridge,Hang,Fogler,Coleridge3,Kramer,Mahan} The semiclassical SdH theory, in fact, ignores the localization and works very well in explaining the low-field magneto-oscillations. \cite{Coleridge} It is expected that the influences of the quantum localization gradually build up with increasing $B$, leading to the breakdown of Eq. (1) before \cite{Chen,Hang,Coleridge2}
\begin{eqnarray}
\Delta \rho _{xx} (B,T) \geq \rho_{xx}(B=0)
\end{eqnarray}
and/or local minimum points of $\rho_{xx} \rightarrow 0$ with decreasing $T$. However, recent experimental studies showed that both Eqs. (2) and (3) can hold true even as $ \rho_{xx} \rightarrow 0$ \cite{Chen,Hang}, and the interests in the applicable range of the semiclassical approaches are revitalized \cite{Piot,Martin}. To understand the transport properties at intermediate magnetic fields, therefore, it is important to investigate the properties of both the semiclassical transport and quantum localization leading to the mobility gap. \cite{Coleridge,Coleridge2,Coleridge3,Fogler,Champel}

To further understand the low-field IQHE, in this paper we present a magneto-transport study on the 2DES in an AlGaAs/GaAs heterostructure. In addition to the extension of Eqs. (2) and (3), \cite{Chen,Hang} we found semiclassical behaviors can coexist with the activation law, which indicates the quantum localization. The value of $\Delta E$, in fact, supports that the mobility gap is due to the electronic states in the Landau-band tails \cite{Oswald1} while the semiclassical behaviors may be attributed to some states away from the tails. Therefore, we shall consider different types of electronic states to understand the survival of semiclassical formula under the quantum localization which leads to the mobility gap. With further increase in magnetic field, we found that the semiclassical LK formula remains valid even as the oscillating amplitude is so large that corrections to the Dingle factor could be attributed to the quantum diffusion \cite{Coleridge3}. 

In the following report, the observations of magneto-oscillations and the analysis based on Eqs. (2) and (3) are presented in Sec. II A. We demonstrate in Sec. II B the existence of a mobility gap by thermally activated conductivity measurements. The mobility gap, which is shown to be related to localization under Landau quantization, can coexist with semiclassical behaviors. Based on our experimental outcome, a possible picture for the crossover from semiclassical transport to the quantum Hall effect is presented in Sec. III. In Sec. IV, we discuss the pre-exponential factor in the thermal activation law as well as the crossover from the Dingle factor to the quantum diffusion model. Conclusions are given in Sec. V. 

\section{EXPERIMENT AND ANALYSIS}
\section*{A. Magneto-oscillations at low magnetic fields}
The sample used for our study is a modulation-doped AlGaAs/GaAs heterostructure grown by molecular beam epitaxy. The 2DES under study resides in the GaAs side of the heterojunction. The 2D channel was followed by a 20 nm spacer layer of Al$_{0.28}$Ga$_{0.72}$As, a 40 nm layer of graded Al$_{x}$Ga$_{1-x}$As (with $x$ from 0.28 to 0) doped with Si at $1 \times 10 ^{18}$ cm$^{-3}$, and a 12 nm GaAs cap layer doped at $1 \times 10 ^{18}$ cm$^{-3}$. The sample was made into a Hall pattern of 0.4 mm width with voltage probes spaced 1 mm apart. Magnetotransport measurements were done with a 14 Tesla superconducting magnet and a He$^{4}$ refrigerator. Magneto-resistance under reversed current is measured to eliminate the thermal voltage. 

The curves of $\rho _{xx}$ at different temperatures in the field region $B$ = 0 - 2.6 Tesla are shown in Fig. 1. At low magnetic fields, the 2DES behaves classically so that $\rho _{xx}$ remains constant. Magneto-oscillations can be observed in  $\rho _{xx}$ as we gradually increase the perpendicular magnetic field $B$. From the oscillating period with respect to $1/B$, the carrier concentration $n = 4.7 \times 10 ^{11}$ cm$^{-2}$ can be obtained. The scattering mobility $\mu _{c}$ obtained by $\rho _{xx}(B=0) = 1/ne \mu _{c}$ is $5.6 \times 10 ^{5}$ cm$^{2}$/Vs.

The carrier effective mass is a quantity that can be quantitatively deduced from the semiclassical SdH theory. It is well-established that the effective mass $m ^{\ast}$ in 2D GaAs electron gases is 0.067 $m _{0}$. Thus, in order to probe the validity range of the SdH theory, we can investigate Eqs. (1) - (3) under the expected effective mass. We note that as $X$ is large enough such that $X/sinhX \sim 2X exp(-X)$ in Eq. (2), the reduced form can be expressed by
\begin{eqnarray}
ln[ \Delta \rho _{xx} (B,T)/T] \sim  C - 2 \pi ^{2} k _{B} m ^{\ast} T/ \hbar eB,
\end{eqnarray}
where $C = ln[2 \pi ^{2} k_{B} m^{\ast}F(B)/\hbar eB]$ is a parameter independent of $T$. Therefore, we can first probe Eq. (2) by checking whether the slope of $ln( \Delta \rho _{xx}/T) -T$ yields the effective mass $m ^{\ast}$ close to the expected value. In our study, the effective mass is about 0.067 $m _{0}$ from the slope as $B < 1.3$ Tesla, where $X$ is large enough to validate Eq. (6). For example, as shown in the inset to Fig. 2, we have $m ^{\ast} \sim 0.069$ $m _{0}$ from the slope of $ln (\Delta \rho _{xx}/T)-T$ at $B$=0.805 Tesla. To exactly examine Eq. (2) at a specific $B$, we can check $\Delta \rho _{xx} = F(B)X/sinhX   \propto X/sinhX$, the expected LK factor, with $m ^{\ast} = 0.067$ $m _{0}$. The dashed line in Fig. 2 shows the fitting at $B$ = 0.805 Tesla. The good fitting in our study justified the extraction of the $T$-independent factor $F(B)$. To further examine the LK factor for all resolved magneto-oscillations, we comprehensively compare $\Delta \rho _{xx}/F(B)$ and $X/sinhX$ in Fig. 3. As can be seen in Fig. 3, the ratio of $\Delta \rho _{xx}/F(B)$ collapses to $X/sinhX$ very well, indicating that the LK formula holds true in our study. Even as $B>1$ Tesla, the consistency still retains although the expected criterion for breakdown, i.e. Eq. (5) becomes valid. In addition, the collapse in Fig. 3 indicates the validity of the LK formula when Eq. (6) fails as $X<1$. Therefore, the semiclassical LK formula is applicable even when Eq. (5) holds true, which is consistent with previous reports. \cite{Chen,Hang} 

The factor $F(B)$ is usually taken as the standard Dingle factor at low magnetic fields. Eq. (3) can be rearranged as
\begin{eqnarray}
ln[F(B)/\rho _{xx} (B=0) ] = ln(4c)- \pi / \mu _{q} B.
\end{eqnarray} 
As shown in Fig. 4, $ln[F(B)/\rho _{xx} (B=0) ]$ is linear with respect to $1/B$ as $B<1.25$ Tesla. The quantum mobility $ \mu _{q}$ can be extracted from the slope of  $ln[F(B)/\rho _{xx} (B=0) ]$ with respect to $1/B$, yielding $ \mu _{q} = 3.5 \times 10 ^{4}$ cm$^{2}$/Vs. The constant $c = 1.2$ is deduced from the intercept. As $B > 1.25$ Tesla, we can see the deviation to Eq. (7) while the LK formula is still applicable. Our study reveals that the LK formula can hold true alone even when we shall consider the corrections to SdH theory to refine Eq. (3). This is going to be discussed in Sec. III.

\section*{B. The mobility gap}
The formation of the mobility gap is expected under the quantum localization, which is important to the IQHE under the high-field Landau quantization. Figure 5 (a) shows the expected density of states in the spin-degenerate IQHE under a perpendicular magnetic field $B$, where the cyclotron gap $ \hbar eB/m ^{\ast}$ separates the centers of adjacent Landau bands. At high $B$, all the electrons are localized except those in the white region of the width $\Gamma _{e}$ near the center of each band. The localized electrons are irrelevant to the conductivity, so an excitation for the change of $\rho _{xx}$ must overcome a mobility gap $\Delta E$ to modify the distribution of the conducting electrons. At low temperatures and/or large effective size, the equation
\begin{eqnarray}
\Delta E \sim \hbar eB/m ^{\ast}
\end{eqnarray}
is expected because $\Gamma _{e}$ is very small. At suitable temperature range where Eq. (4) is valid, we have
\begin{eqnarray}
ln \rho _{xx,min} = ln \rho _{0} - \Delta E/ 2 k _{B} T
\end{eqnarray}     
at the minimum points of $\rho _{xx}$ in $B$. When the spin-splitting is unresolved, such points correspond to the even filling factors, and the Fermi energy $E _{F}$ is expected to be located near the middle point between adjacent Landau bands in Fig. 5 (a). 

The mobility gap at $B=0.74-2.43$ Tesla is obtained from the fitting according to Eq. (9) at the even filling factors from 8 to 26 under suitable temperature range, as shown in the inset to Fig. 6. The good linear fitting of $ln \rho _{xx,min}$ with respect to $T ^{-1}$ indicates the existence of the mobility gap in such a magnetic-field region, where the semiclassical LK formula or Eq. (2) holds true as shown in Fig. 3. Figure 6 shows the obtained $\Delta E$ as a function of the magnetic field. We can see that the gap $\Delta E$ is linear in $B$ with the slope 1.6 meV/Tesla, which is 92$\%$ of $ \hbar e/m ^{\ast}$ as estimated by Eq. (8) with $m^{\ast} = 0.067$ $m _{0}$, the well-established value in the 2D GaAs electron gases. Hence the slope of  $E - B$ provides the quantitative evidence for the mobility gap to be attributed to Landau quantization. In our study, such a gap exists while the LK formula applies. Besides, between $B = 0.74 \sim 1.25$ Tesla, the standard Dingle factor given by Eq. (3) can also be fitted well, which is accompanied by the formation of the gap. Therefore, we found the coexistence of the mobility gap, the semiclassical LK formula, and the Dingle factor in the low-field IQHE before the ultimate deviation to semiclassical descriptions in strong magnetic fields.

\section{FROM SEMICLASSICAL TRANSPORT TO QUANTUM HALL EFFECT}
In the crossover from semiclassical transport to the IQHE, the semiclassical LK formula and Dingle factor appearing in the SdH transport theory can survive when the formation of the mobility gap indicates the importance of the quantum localization. To explain the unexpected coexistence in the crossover, we note that the semiclassical approach works best when Landau quantization modulates the density of states without inducing noticeable quantum localization at low enough $B$ in most realistic 2DESs. It is essential that when the localization effects gradually take over with increasing $B$, the localization length is not the same for all states. Because the localization length increases rapidly near each Landau-band center \cite{Wei,Fogler}, the onset of the localization should emerge from the tails of each Landau band. In the intermediate magnetic-field region, as shown in Fig. 5 (b) \cite{Oswald1}, it is possible that only tails of Landau bands, shown as black regions, are fully occupied by localized electrons. Meanwhile, the mobility gap is reduced because of the broadening of the width  $\Gamma _{e}$ due to the insufficient localization, shown as the white regions in which the electrons are not all localized. Moreover, the survival of the semiclassical LK formula under the apparent formation of the mobility gap could be attributed to a distribution of the semiclassical conducting electrons away from the Landau-level tails \cite{Oswald1}. The non-zero intercept in Fig. 6, in fact, indicates $ \Delta E \sim \hbar eB/m ^{\ast} - \Gamma _{e}$ and provides the quantitative width value $\Gamma _{e} = 0.51$ meV for the white regions in Fig. 5 (b). In our study, the value of $\Gamma _{e}$ is in good agreement with the usual broadening measure $\hbar/ \tau _{q}$ = 0.50 meV, where the quantum lifetime $\tau_{q} =\mu _{q} m ^{\ast}/e$.

With further increase in magnetic field, the oscillation amplitude becomes comparable to the zero-field resistivity value. The longitudinal resistivity goes to zero at its minima, which is regarded as a characteristic of IQHE. Here we start to see the deviation of the semiclassical approach in the Dingle plot as $B>1.25$ Tesla. However, as Fig. 3 shows, the LK formula remains surprisingly valid with the existence of thermal activated conductivity due to quantum localization. Therefore we provide firm evidence that the LK formula is more robust than the Dingle factor with respect to quantum localization. As shown in the inset to Fig. 4, we found that $F(B) \propto B$ when there exists the deviation to the Dingle factor. We note that Coleridge \cite{Coleridge3} has derived the equation
\begin{eqnarray}
\rho _{xx} ^{peak} = \rho _{xx} (B = 0) \mu_{q} B,           
\end{eqnarray}
for the deviation by considering the quantum diffusion effects. Here $\rho_{xx} ^{peak}$ denotes the peak value of $\rho _{xx}$ at low temperatures. Because the minimum of $\rho _{xx}$ approaches zero with decreasing $T$ when Eq. (3) fails, we can expect
\begin{eqnarray}
2F(B) \sim \rho _{xx} ^{peak} = \rho _{xx} (B=0) \mu _{q} B \propto B
\end{eqnarray}
as $T \rightarrow 0$ such that $X/sinhX \rightarrow 1$. Therefore, the experimental result $F(B) \propto B$ is consistent with Eq. (10). Besides, the slope of $2F(B)- B$ in our study is only 10 $\%$ larger than the product $ \rho _{xx} (B = 0) \mu_{q}$, which provides the quantitative evidence to the validity of Eq. (10). In view of the deviation to the Dingle factor shown in Fig. 4, our results suggest the importance of the quantum diffusion to the corrections of the Dingle factor in the departure from the semiclassical transport. 

\section{DISCUSSION}

Conventional description for the quantum magneto-oscillations is based on the semiclassical model, in which the discrete zero-field 2D density of states are evolved into broadened Landau levels. Note that the quantum localization plays no role in this approach. When the modulation to the density of states is not large at low magnetic fields, the oscillation amplitude is small and the analytic formulas given by Eqs. (1) - (3) can be derived from the semiclassical SdH theory. As the oscillation amplitude gradually increases with increasing $B$, this description is expected to be invalid. However, our analysis shows that both the LK factor and the Dingle factor hold true to a field strength larger than expected. Hence the applicable range of the semiclassical description can be extended beyond the region implied by the conventional derivation. \cite{Chen,Hang,Martin} From a practical point of view, this suggests that the condition for magneto-oscillation analysis may be less stringent than general belief. Because of the positivity of $\rho _{xx}$, Eq. (1) cannot hold true when Eq. (5) is valid. It has been proposed that Eq. (1) should be refined by incorporating the positive magneto-resistance background for the extension of Eqs. (2) and (3). \cite{Chen,Hang} Consistent with this point of view, the dashed line in the inset to Fig. 1 shows such a background at $T = 1.9$ K. In addition to the non-oscillatory background, there also exists distortion on the oscillating factor $cos[\pi(\nu-1)]$ in Eq. (1) as the minimum of $\rho_{xx}$ approaches zero in our study. Different mechanisms have been discussed to understand the deviations on the semiclassical SdH theory given by Eqs. (1) - (3) as the magnetic field is increased. \cite{Coleridge,Coleridge2,Champel,Coleridge3,Fogler}

In addition, we found the Dingle plot in Fig. 4 yields the constant $c = 1.2 \sim 1$, the expected value for the Dingle factor. The small deviation of the constant $c$ to the expected value \cite{Chen,Hang,Cho}, in fact, is important to the crossover from the Dingle factor to Eq. (11). A direct crossover can occur when 
\begin{eqnarray}
F(B) = 4c \rho _{xx} (B=0)exp(-\pi / \mu _{q} B) = \frac{1}{2} \mu _{q}\rho _{xx} (B=0)B.     
\end{eqnarray}
If c = 1 exactly, the second equality fails and there should be no direct crossover. The direct crossover in our study is evidenced by the fact that both Eqs. (3) and (11) apply between $B = 0.74 - 1.25$ Tesla. As the magnetic field is additionally increased in the crossover, the factor $F(B)$ approaches Eq. (11) while it departs from Eq. (3). We note that the current carried by the edge channels \cite{Halperlin,Oswald2}, the voltage drop near the current injection points \cite{Kramer2}, and the shape of Landau bands \cite{Coleridge3} are all important to the details of magneto-tranport. Hence more studies are necessary to further understand the criteria for the validities of Eqs. (2) and (3).   

It is known that the mobility gap $\Delta E$ in Eq. (4) can deviate from the cyclotron gap under the variations on the positions of extended states \cite{Khmel'nitzkii}, electron-electron interaction \cite{enhancement}, and the existence of quasiparticles \cite{Champel}. In our study, the value of $\Delta E$ is close to the expected cyclotron gap, and the energy $\hbar/ \tau _{q} = \hbar e / \mu _{q} m ^{\ast}$ determined by the Dingle factor or by the quantum diffusion model is close to the zero-field intercept of the solid line in Fig. 6.  Such an intercept, in fact, indicates that the mobility gap disappears as the cyclotron energy becomes close to the broadening measure $\hbar/ \tau _{q}$ with decreasing $B$, as reported in Ref. \cite{Du}.    

Finally, we discuss the pre-exponential factor in the equation for thermal activation. By using the Hall resistivity $ \rho _{xy} = h/\nu e ^{2}$ for quantum Hall state of filling factor $\nu$, Eq. (4) can be converted into the equation for longitudinal conductivity $\sigma _{xx}$. We then have
\begin{eqnarray} 
\sigma _{xx} \underline{\sim} \sigma _{0} exp(- \Delta E/2k_{B}T)     
\end{eqnarray}
with the prefactor $\sigma _{0} =  \rho _{0,\nu}/(h/ \nu e^{2})$. Different experimental results on $\sigma _{0}$ have been reported. \cite{Clark,Usher,Shih} For GaAs/AlGaAs heterostructures, Clark et al. \cite{Clark} reported $\sigma _{0}$ value close to $e ^{2}/h$ at $\nu = 2$, 4 and 6, while $\sigma _{0}$ value $\sim 2 e^{2}/h$ at $\nu = 2$ have been obtained by Usher et al. \cite{Usher}. In our study, the pre-exponential factor is found to be $ \sim 2 e ^{2}/h$ at $B = 0.74$ Tesla. As the magnetic field increases, we found $\sigma _{0}$ gradually drops to $ \sim e ^{2}/h$. A similar result has also been reported for a SiGe/Si quantum well, in which $\sigma _{0}$ is found to be $1.2 e ^{2}/h \sim 2 e^{2}/h$ between $\nu = 4$ and $\nu = 16$. \cite{Shih}  

\section{CONCLUSION}
In conclusion, we have performed a magnetotransport measurement on a modulation-doped AlGaAs/GaAs heterostructure. The behaviors from semiclassical transport to quantum localization are investigated. The extended applicable range of the LK formula is carefully demonstrated. The activation energy study indicates that the origin of the mobility gap is attributed to the localization under Landau quantization. In addition, we found the coexistence of the LK formula and the Dingle factor with the mobility gap in the crossover. The survival of the semiclassical LK formula under the mobility gap can be attributed to a distribution of conducting electrons away from the Landau band tails. While the LK formula remains valid with large oscillating amplitudes, we show that the departure from the Dingle factor implies a smooth transition to the quantum diffusion model. Our results suggest that different types of electronic states should be taken into account to understand the coexistence of the semiclassical transport and quantum localization. 

\section*{ACKNOWLEDGMENTS}
This work is supported by the National Science Council of the Republic of China under grant no: NSC 96-2112-M-110-006-MY3. We thank S. H. Lo and J. C. Hsiao for experimental help. D. R. Hang acknowledges financial support from Aim for the Top University Plan of National Sun Yat-sen University, Taiwan. 


Figure caption \newline
Figure 1: The longitudinal resistivity as a function of magnetic field for different temperatures (from top to bottom are 1.9, 2.2, 2.7, 3.1, and 4.3 K). The thermal offset has been removed. The dashed line in the inset shows the nonoscillatory background at $T$= 1.9 K obtained by averaging the magneto-oscillations. \newline
Figure 2: The inset shows the plot $ln( \Delta \rho _{xx} / T )$ versus $T$ at $B$=0.805 Tesla. According to the LK factor, the carrier effective mass can be extracted from the slope. The obtained value is about 0.069 $m _{0}$ which is close to the expected value. Accordingly, the curve of $F(B)(X/sinhX)$ and the experimental points of $\rho _{xx}$ at $B$=0.805 Tesla can be displayed. At this magnetic field, Eq. (2) is valid because the experimental points are all near the curve $F(B)(X/sinhX)$. \newline
Figure 3: To check the LK formula for all resolved magneto-oscillations, we comprehensively plot $\Delta \rho _{xx}/F(B)$ with respect to $X$ for various fixed temperatures. The symbols squares, circles, up triangles, down triangles and diamonds are for the points at $T$ = 1.9 K, 2.2 K, 2.7 K, 3.1 K and 4.3 K, respectively. The range of magnetic field extends to 2.17 Tesla. The numerical evaluation of $X/sinhX$ as a function of $X$ is shown as the solid line. The curve of $\Delta \rho _{xx}/F(B)$ collapses well into a single curve $X/sinhX$ with respect to the parameter $X$. \newline
Figure 4: $ln[ F(B) / \rho _{xx} (B=0) ]$ as a function of inverse magnetic field, from which the quantum mobility can be obtained. The inset shows $2F(B)$ as a function of magnetic field and the fitting to Eq. (11). \newline
Figure 5: (a) Density of states in Landau bands for spin-degenerate 2DES in a perpendicular magnetic field. At high $B$, only electrons in the white regions of width  e near the center of each band are delocalized. 
(b) For the intermediate field strength, only electrons in the tails, shown as black regions, are well-localized. The mobility gap is consequently reduced. The dashed line shows the schematic distribution of electrons responsible for the semiclassical transport. \newline
Figure 6: The activation energies as a function of magnetic field. The solid line is the linear fit to the data. The zero-field intercept is indicated by the dashed line. The inset shows the Arrhenius plot of $\rho _{xx,min}$ with linear fits (from top $B$ = 0.745, 0.808, 0.881, 0.969, 1.077, 1.212, 1.389, 1.619, 1.944, and 2.429 Tesla).

\end{document}